\documentclass[12pt]{iopart}

\usepackage{graphicx}

\begin{document}

\title{Derivation of sensitivity of a Geiger mode APDs detector from a given efficiency for QKD experiments}

\author{Kiyotaka Hammura and David Williams}

\address{Hitachi Cambridge Laboratory, Cavendish Laboratory, J J Thomson Avenue, Cambridge CB3 0HE, United Kingdom}
\ead{kh401@cam.ac.uk}
\begin{abstract}
The detection sensitivity ($DS$) of the commercial single-photon-receiver based on InGaAs gate-mode avalanche photodiode is estimated. 
Instalment of a digital-blanking-system (DBS) to reduce dark current makes the difference between $DS$, which is an efficiency of the detector during its open-gate/active state, and the total/overall detection efficiency ($DE$).  
By numerical simulations, it is found that the average number of light-pulses, blanked by DBS, following a registered pulse is 0.333.
$DS$ is estimated at 0.216, which can be used for estimating $DE$ for an arbitrary photon arriving rate and a gating frequency of the receiver.

\end{abstract}

\noindent{\it Keywords}: Quantum Cryptography, Detection sensitivity and efficiency
\maketitle

\section{Introduction}
The emergence of a commercial single-photon-receiver (SPR) would help reduce development time to integrate quantum key distribution (QKD) system.  
Since QKD\cite{Bennett1984} was firstly experimentally demonstrated\cite{Bennett1992}, tremendous efforts have continued towards QKD's further performance improvements\cite{Gisin2002}.
Many efforts have been dedicated to the development of two essential devices which constitutes QKD, i.e., single-photon-emitters\cite{Xiulai2004,Xiulai2007,Xiulai2008} and SPRs\cite{Bethune2004,Lacaita1996,Zappa1994,Cova1996,Stucki2002,Dautet1993,Golovin2004,Vasile1998,Kim1999,Yoshizawa2003,Kang2003}. 
For current QKD systems under development, almost all of the performances depend on those of SPRs. 
This is because the single-photon-emitters, the other essential device, have normally been implemented by attenuated commercial lasers.
The performances of them have been completely characterised already, meaning that there is no room to improve.     
The commercial SPR of Princeton Lightwave\cite{PLI-AGD-SC-Rx}, for example, is a highly integrated one.
This SPR features not only high detection efficiencies, $DEs$, at telecom wavelengths and low dark count probabilities but also a digital-blanking system (DBS) to dramatically reduce the effect of afterpulsing on dark count.
  
It should be noted, however, that the given $DE$ does not clarify about some intrinsic figure-of-merit of the SPR concerning its sensitivity to incident single-photons.
We should have a figure-of-merit of how sensitive the SPR in its open-gate state is (which will be designated as a detection sensitivity, $DS$, hereafter) while the given $DE$ is simply a ratio of the number of registration to that of all incident light-pulses.
The verbal definitions of these two figures follow:
\begin{eqnarray}
	DE & \equiv &\frac{\textit{How many light-pulses are registered}}{\textit{How many light-pulses hit the detector}}\,,\label{def_de}\\
	DS & \equiv &\frac{\textit{How many light-pulses are registered}}{\textit{How many light-pulses hit the detector in open-gate status}}\label{def_ds}.
\end{eqnarray}

Thanks to the DBS, a designated number of bias pulses for gating the detector are neglected once a registration occurs.  
In this paper, $DS$ of the commercial SPR\cite{PLI-AGD-SC-Rx} is estimated.
$DS$ is helpful to estimate the photon registration number for various operation conditions while $DE$ is helpful only for a set of conditions to give the $DE$. 
 
$DS$ should be defined as a ratio of the number of registrations to that of light-pules which hit the detector in open-gate status (Eq.(\ref{def_ds})).
Such number of light-pulses is to be estimated by subtracting the number of light-pulses blanked by DBS, $N_{B}$, from all incident ones.
The $N_{B}$ will be numerically estimated as below.

\section{Method}
\begin{figure}[htbp]
	\centering
		\includegraphics[height=3in]{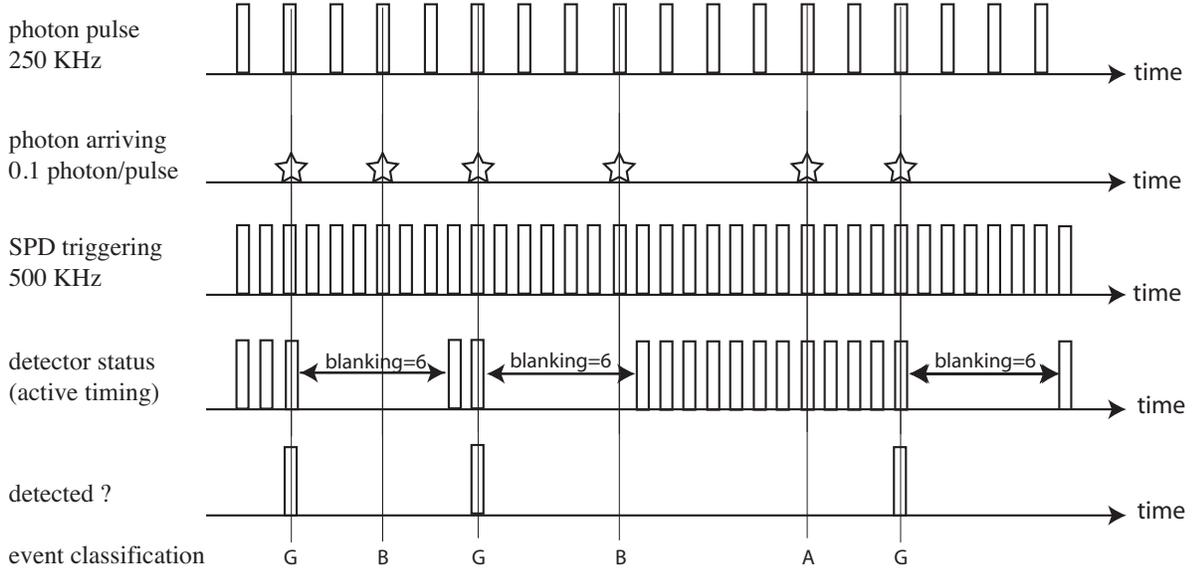}
	\caption{$N_{A}$, $N_{B}$, and $N_{G}$ in the text refer to the numbers per second of incidents denoted by $A$, $B$, and $G$.}
	\label{fig:NB_simulation_m2}
\end{figure}

$DS$ can be derived using one experimental data and the number of light-pulses blanked by DBS.
Eq.(\ref{def_ds}) is expressed as,
\begin{eqnarray}
DS & = & \frac{N_{G}}{N_{A}+N_{G}}\\
& = & \frac{N_{G}}{({N_{A}+N_{G}+N_{B}})-N_{B}}\,,\label{denominator}
\end{eqnarray}
where $N_{G}$ and $N_{A}$ designate the numbers per second of how many light-pulses are registered, how many light-pulses hit the detector in gate-open status but are not registered due to detector's poor sensitiveness as a whole, respectively.
The relations among $N_{G}$,$N_{A}$, and $N_{B}$ are illustrated in FIG \ref{fig:NB_simulation_m2}.
We use an experimental data given by the manufacturer\cite{princetonlightwave} for $N_{G}$ in the numerator in Eq.(\ref{denominator}).

$N_{B}$ is calculated under the following experimental conditions:
the light-pulse arriving rate, $N_{T}=250\times 10^3$ s$^{-1}$, the mean number of photons per pulse, $\mu =0.1$, the triggering rate of the detector, $N_{TR}=500\times 10^3$ s$^{-1}$, the number of gatings blanked by DBS, $B_{L}=6$.  
The simulation is conducted according to the following 3 steps. All figures below are those for one second.
\begin{enumerate}   
\item{Distribute 25000 ($=N_{T} \cdot \mu\times 1\,\,\rm{s}\equiv N_{0}$) incident single-photons randomly over 250000 ($=N_{T} \times 1\,\,\rm{s}$) time-bins. The specific procedure is to take a random number between 1 to 250000, for each of 25000 figures.  Then, arrange the 25000 figures in ascending order of their associated random numbers.  The random numbers mean the occupied time-bins.} 
\item{Estimate how many light-pulses in average, designated as $N_{B,AVG}$, are found sitting in the consecutive 3 blanked bins following the time-bin which is actually registered.  
To do this, choose randomly an occupied time-bin and count up the number of photon-pulses in the blanked bins,  resulting in $N_{B,AVG}$. Repeat this procedure until $N_{B,AVG}$ reaches some asymptotic number.
The reason why the number of blanked bins is not 6 ($=B_{L}$) but 3 comes from the fact that the triggering rate, $N_{TR}$, is twice faster than the light-pulse arriving rate, $N_{T}$.  Blanking 6 time-bins only means blanking 3 pulses.
} 
\item{$N_{B}$ is given by multiplying an experimental data value of $N_{G}$ by $N_{B,AVG}$.}
\end{enumerate}

\section{Result}
Numerical simulations of $N_{B,AVG}$ results in 0.333 s$^{-1}$. 
With Eq.(\ref{denominator}), $DS$ is estimated at 0.216 as follows:
\begin{eqnarray}
	DS 	& = & \frac{N_{G}}{(N_{A}+N_{B}+N_{G})-N_{G}\times N_{B,AVG}}\\
		& = & \frac{5033}{25000-5033\times 0.333}=0.216\,\,,
\end{eqnarray}
where $N_{G}=5033$ s$^{-1}$\cite{princetonlightwave} and $N_{A}+N_{B}+N_{G} \equiv N_{0}$.

$DE$ for an arbitrary combination of the light-pulse arriving rate, the triggering rate, the mean photon number per pulse, and DBS number, $(N_{T},N_{TR},\mu, B)$, can be estimated using the constant value of $DS$.
The numerator in the definition of $DE$ in Eq.(\ref{def_de}) is decomposed as
\begin{equation}
	DE\equiv \frac{N_{p}\times DS}{N_{0}}\,\,.
\end{equation}	
, where $N_{p}$ designates the quantity of how many photons {\it could be} registered at the maximum.
For $N_{0}$,
\begin{equation}
	N_{0}=\left\{
		\begin{array}{@{\,}ll}
			N_{TR}\times \mu & \,\,\,(N_{T}\geq N_{TR})\\
			N_{T}\times \mu & \,\,\,(N_{T}\leq N_{TR})
	    \end{array}
	\right. 
\end{equation}	
For $N_{p}$, numerical simulations for $\mu=0.1$ and $B_{L}=6$ will be done as follows: Distribute incident light-pulses, $N_{dist}$ (=$N_{0}$), randomly over the time-bins, $N_{bin}$ (=$N_{TR}$), resulting in identifying occupied bins. Then, count up how many occupied bins can be chosen at the maximum so that if you choose an occupied bin it is not allowed to count up any occupied bins sitting in 6 time-bins following the chosen occupied bin to meet $B_{L}=6$.  

\begin{figure}[htbp]
	\centering
		\includegraphics[height=3in]{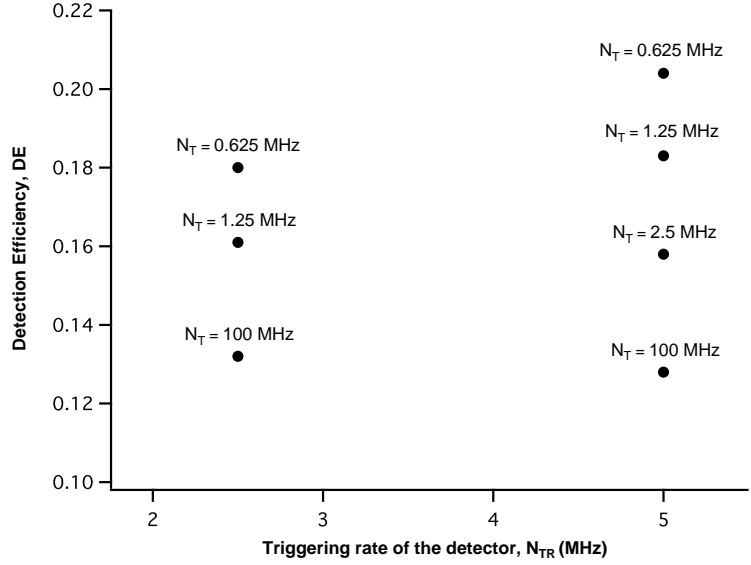}
	\caption{Plots of Detection Efficiency, $DE$, for several combinations of light-pulse arriving rate, $N_{T}$, and triggering rate of the detector, $N_{TR}$, under $\mu=0.1$ and $B_{L}=6$. $DE$ increases with increasing $N_{TR}$ as long as $N_{T}\leq N_{TR}$, reflecting decreasing in the probability of a distributed photon being found sitting in the blanked time-bins with increasing total time-bins.  $DE$ increases with decreasing $N_{T}$ for either $N_{TR}$, reflecting decreasing in the probability of a distributed photon being blanked due to the sparse distribution.}
	\label{DE_plot}
\end{figure}

Simulated results of $N_{p}$ and $DE$ for several sets of $(N_{T},N_{TR})$ are summarised in Table \ref{DEs} and Figure \ref{DE_plot}.

For a fixed triggering rate of the detector, $N_{p}$ do not decrease so rapidly as $N_{0}$, leading to the increase in $DE$. 

\begin{table}
	\caption{\label{DEs} Results of the numerical simulation of $N_{p}$ and the resultant $DE$ for several specific combinations of the photon arrival rate and the detector triggering rate. The mean photon number and the DBS number are fixed at 0.1 and 6, respectively. }
\begin{center}
\begin{tabular}{c|ccc|c||c}
($N_{T},N_{TR}$) & $N_{0}$ & $N_{bin}$ & $N_{dist}$ & $N_{p}$ & $DE$\\
\hline
(100 MHz,5 MHz) & $500\times 10^{3} s^{-1}$ & $5\times 10^{6} s^{-1}$ & $500\times 10^{3} s^{-1}$ & $296\times 10^{3} s^{-1}$ & 0.128\\
(2.5,5) & $250\times 10^{3}$ & $5\times 10^{6} $ & $250\times 10^{3}$ & $183\times 10^{3}$ & 0.158\\
(1.25,5) & $125\times 10^{3}$ & $5\times 10^{6} $ & $125\times 10^{3}$ & $106\times 10^{3}$ & 0.183\\
(0.625,5) & $62.5\times 10^{3}$ & $5\times 10^{6} $ & $62.5\times 10^{3}$ & $59\times 10^{3}$ & 0.204\\
(100,2.5) & $250\times 10^{3}$ & $2.5\times 10^{6} $ & $250\times 10^{3}$ & $153\times 10^{3}$ & 0.132\\
(1.25,2.5) & $125\times 10^{3}$ & $2.5\times 10^{6} $ & $125\times 10^{3}$ & $93\times 10^{3}$ & 0.161\\
(0.625,2.5) & $62.5\times 10^{3}$ & $2.5\times 10^{6} $ & $62.5\times 10^{3}$ & $52\times 10^{3}$ & 0.180\\
\end{tabular}
\end{center}
\end{table}  

\section{Discussions}
By estimating the effect of blanking on the number of photon detection registration, DS of the detector is derived under the condition that dark count events are all neglected. 

We established a method to derive DS, which is more suitable figure-of-merit to characterise device parameters of SPADs.

The estimation shows that 0.333 light-pulse is in fact blanked by DBS in average.  This correction might look negligible in experiments.  
The importance of the figure, however, does not lie in its magnitude but in that it enables us to derive the significant performance characteristic intrinsic to the specific detector.

The intrinsic $DS$ value enables us to estimate detection efficiencies, $DE$, under arbitrary combinations of operation parameters, {\it i.e.}, $N_{T}, N_{TG}, \mu, B_{L}$. 
For a fixed $N_{T}$, $DE$ increases with increasing $N_{TR}$.
This is because the probability of a distributed light-pulse being sitting in the 6 time-bins (=blanked bins) following a registered pulse decreases with increasing triggering rate as the triggering rate means the number of time-bins.
For a fixed $N_{TG}$, $DE$ increases with decreasing $N_{T}$. 
This is because the probability of a distributed light-pulse being sitting in the blanked bins decreases faster than the decrease in the total number of distributed light-pulses.  
The discrepancy in decreasing speed of the two quantities depends on the number of DBS setting.

In the development of optical communications systems using single photon detectors, it is crucial to know in prior how many photons will be registered under the perfect operation condition.  
$DS$ enables us to do this.  
For example, applying $DS=0.216$ to our experimental configurations: $N_{TR}=5\times 10^6$ s$^{-1}$ and $N_{T}=100\times 10^6$ s$^{-1}$, the total photon count for a single photon stream is estimated at 64800 s$^{-1}$ (=$N_{p}\times DS$). 
This figure is helpful in the respect of system optimization because it gives the maximum number of photon registration achievable in the configurations.   

Throughout all the calculations in this paper including the derivation of $DS$ and the estimation of $DE$, the dark count event is neglected for simplicity.  
The legitimacy of this approximation is able to be checked in experiments using a single photon emitter implemented by an attenuated laser.
 
Even if the general characteristic of $DE$ associated with the specific detector under various operation conditions would be successfully verified experimentally, a potential future research is to evaluate $DS$ and $DE$ more precisely with more elaborate simulation model including dark count events.


\begin{thebibliography}{9}
\bibitem{princetonlightwave}
	Figures 5033 is given through private communications with the manufacture of the detector (Princeton Lightwave Inc.).
\end{thebibliography}
\end{document}